# THERMAL PROPERTIES OF HIGH POWER LASER BARS INVESTIGATED BY SPATIALLY RESOLVED THERMOREFLECTANCE SPECTROSCOPY

*Dorota Pierścińska[1], Kamil Pierściński[1], Maciej Bugajski[1], Jens W. Tomm[2]*

[1]Institute of Electron Technology, Al. Lotnikow 32/46, 02-668 Warsaw, Poland
[2]Max-Born-Institut, Max-Born-Str. 2 A, 12489 Berlin, Germany

## ABSTRACT

In this work we present results of the analysis of thermal properties of high-power laser bars obtained by spatially resolved thermoreflectance (TR) spectroscopy.

Thermoreflectance is a modulation technique relying on periodic facet temperature modulation induced by pulsed current supply of the laser. The periodic temperature change of the laser induces variation of the refractive index and consequently modulates probe beam reflectivity. The technique has a spatial resolution of about ~1 µm and can be used for temperature mapping over 300 µm × 300 µm area. Information obtained in these experiments provide an insight into thermal processes occurring at devices' facets and consequently lead to increased reliability and substantially longer lifetimes of such structures.

## 1. INTRODUCTION

High – power optoelectronic devices such as diode lasers, diode laser arrays or bars have steadily grown in power and decreased in price per watt. These devices offer a wide range of applications, e.g. as efficient sources for pumping solid-state lasers[1,2] (with a high spatial and spectral quality), as direct tools for material processing[3,4] and in medicine[5].

However, limitations regarding device lifetime and reliability still limit their access to all potential application fields. Reabsorption of the laser radiation[6] and catastrophic optical damage (COD)[7] are among the mechanisms which cause laser bar degradation and consequently limit emitted optical power and device lifetime.

Several thermometric techniques such as thermoreflectance[8], micro – Raman spectroscopy[9], and micro-photoluminescence[10] are used to determine temperature distribution in the semiconductor laser. Information obtained in these experiments provide an insight into thermal processes occurring at devices' facets and consequently lead to increased reliability and substantially longer lifetimes of such structures.

In this work we use thermoreflectance (TR) spectroscopy to obtain temperature distribution over the facet of individual emitters in high power laser bars. The high temperatures, in excess of 200°C have been observed for devices undergoing the failure. The increase of facet temperature with operation time for these emitters suggests thermal runaway model of COD. The damage to the facet has been confirmed by scanning electron microscope (SEM) investigation. Our investigation confirm that very high facet temperatures exist for operation currents between the failure of individual emitters and the total device failure.

## 2. THERMOREFLECTANCE TECHNIQUE

### 2.1. Basic concepts

The heating of the facet in high – power laser bars and its degradation has been studied by means of non – invasive optical technique: thermoreflectance[11,12].

TR is a modulation technique relying on periodic facet temperature modulation induced by pulsed current supply of the laser (repetition rate 50-200 Hz, duty cycle 25-50%). Under these conditions the laser is operating quasi-CW and is subjected to thermal conditions, which are almost similar to genuine CW operation. The spatially resolved TR technique (spatial resolution of about ~ 1μm) provides detailed temperature distribution maps of operating devices.

The relative variation of sample reflectance ΔR/R is linear versus the temperature variation ΔT and a measurement of the sample reflectance allows for the determination of the local temperature increase using the following formula:

$$\Delta T = \left(\frac{1}{R}\frac{\partial R}{\partial T}\right)^{-1}\frac{\Delta R}{R} \equiv \kappa \frac{\Delta R}{R} \qquad (1)$$

Here κ is the TR coefficient which depends on probed material as well as on the wavelength of probe light. Therefore it should be determined experimentally for each investigated device. Here, the TR coefficient has been determined experimentally by means of micro-photoluminescence (PL) measurement. More information about calibration of thermoreflectance by photoluminescence measurements can be found in [11]. The other method which provides more straightforward temperature determination is micro – Raman spectroscopy. The absolute value of temperature can be obtained from the intensity ratio of anti-Stokes and Stokes lines or independently from the temperature dependent shift of these lines positions. This technique is preferred for determination of the value of TR coefficient[13].

2.2. Experiment

In TR measurement devices are mounted on a temperature-controlled heat sink positioned on a high precision $x – y – z$ piezo-translator stage to align laser mirror to the laser probe beam which is reflected from the facet. Using piezoelectrically driven translator stage allows for mapping laser facet temperature with sub – micron resolution. The experimental set-up used in the experiments is shown in Fig. 1. The clear advantage of thermoreflectance over micro-Raman measurements, which also have mapping capability, manifest itself in that TR, contrary to Raman, is a single-wavelength measurement.

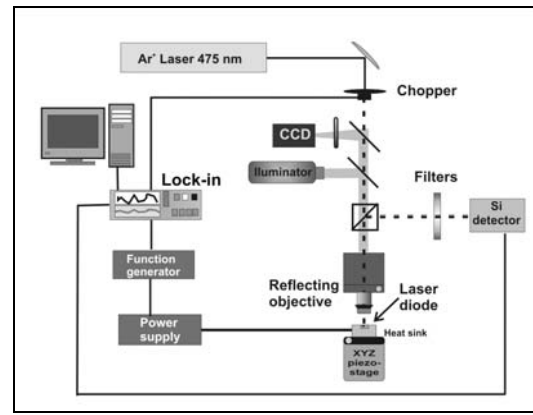

**Fig 1. Experimental set-up for thermoreflectance measurements on laser diode and laser bar mirrors.**

As the probe beam, the 475 nm (2.61 eV) line of an Argon laser was used. The probe beam was focused with a microscope objective to the spot of diameter $\varnothing_{1/e}$ ~ 0.6 μm. The probe beam intensity was about 1 mW. The laser facet was simultaneously visualized by a CCD camera, which allows for accurate probing of specific regions of the laser. The reflected beam was directed onto a Si photodiode detector. The resulting output was fed into a lock-in amplifier. The signal obtained was normalized by the simultaneously measured DC reflectivity component (R in eq. 1) resulting in the information on the relative reflectance change (TR-signal). More details on the experimental technique is given in Ref. 11.

The investigated devices were high power 1 cm laser bars with 19 emitters of 200 μm wide stripes. The individual emitters structures consisted of 7 nm thick $In_{0.06}Ga_{0.86}Al_{0.08}As$ double quantum wells separated by a 8nm barrier embedded in 1μm thick $Ga_{0.7}Al_{0.3}As$ waveguide, surrounded by 1.5 μm thick $Ga_{0.4}Al_{0.6}As$ cladding layers. The facets of investigated laser bars have 120 nm thick $Al_2O_3$ anti-reflection (AR) coating. The laser bars were pulse operated at currents up to 50 A and duty cycle of 20%. The temperature of the device was stabilized at $25^0C$ by water cooled Peltier element.

## 3. RESULTS AND DISCUSSION

For each emitter the temperature line scans starting at the active region towards the substrate were measured. The measurements were repeated three times and the results averaged. The temperature maps (300 μm x 120 μm) were measured only for selected emitters. Fig.2. presents distribution of active region temperatures in laser bar, together with temperature line scans for selected emitters, showing characteristic behaviour. All emitters, except a few, show the facet temperature increase of the order of $30^0C$. Three emitters are substantially hotter, with temperature of emitter no. 8 reaching $200^0C$. Results of consecutive scans for each emitter showed the same values of temperature except for emitter no. 18 which exhibited increased degradation after each scan.

The emitter no. 8 exhibited the highest temperature of the active region among all emitters. The map of temperature distribution for emitter no. 8 is shown in Fig. 3. The map was recorded with 0.5 μm step in x and y direction. Black colour corresponds to no change of temperature and white colour to the highest change of temperature. The facet of this emitter is highly defected, as can be seen in a picture from CCD camera built in the experimental set-up (Fig. 4).

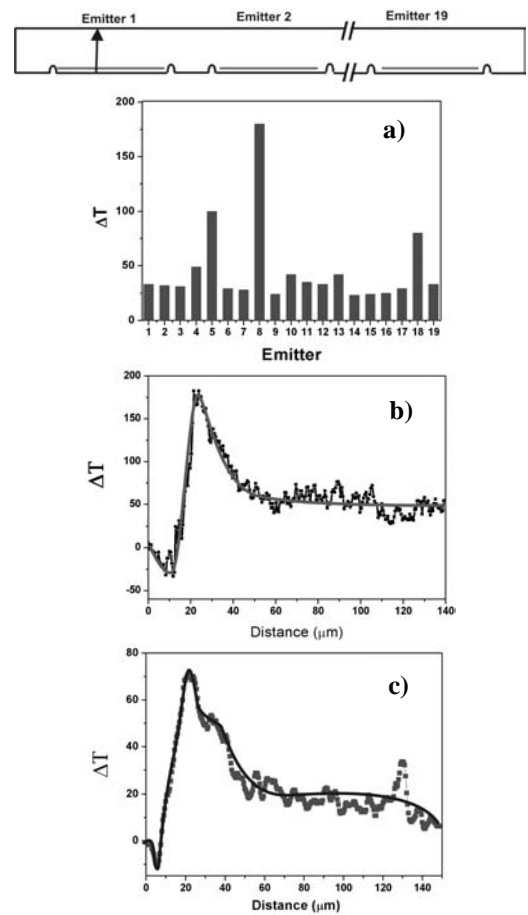

**Fig 2.** Distribution of active region temperatures in laser bar (a) and temperature line scans for selected emitters; (b) – emitter no. 8, (c) – emitter no. 18.

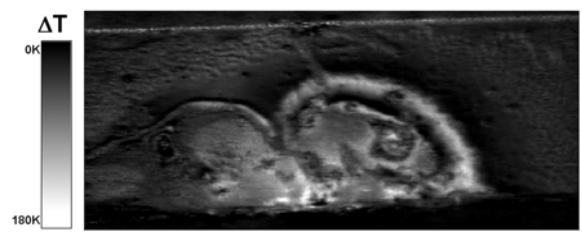

**Fig 3.** Temperature map of the emitter no. 8 at I=50A and at the heat sink temperature of 25°C.

The temperature distribution map has features closely corresponding to CCD camera image and scanning electron microscope (SEM) pictures which evidence damage of this particular emitter AR coating. The normalization of the signal by DC component (R in eq. 1) should account for modified reflectance in the areas of damaged coatings.

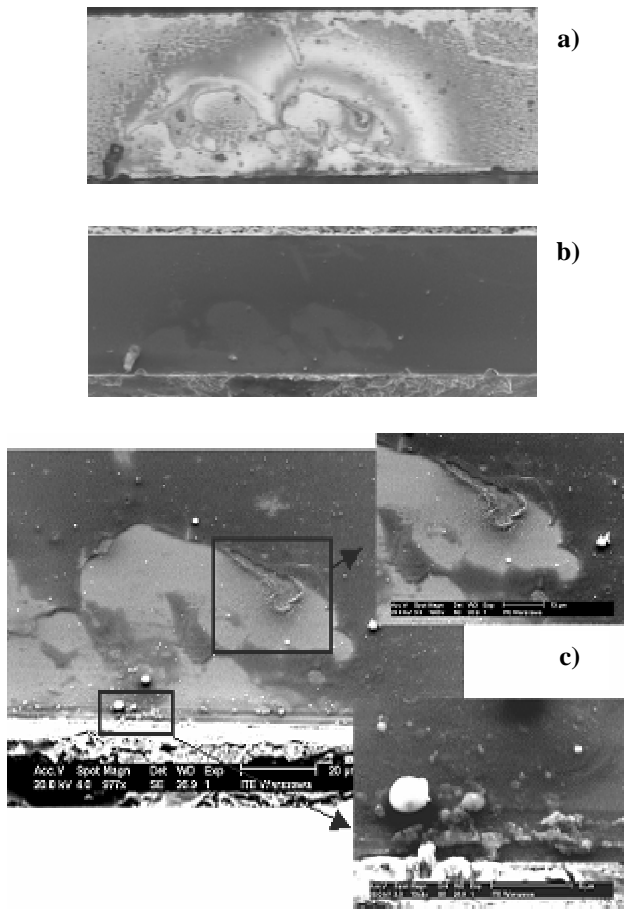

**Fig 4. CCD camera image of the facet of emitter no. 8 - (a), and SEM photo of emitter no. 8 showing damaged antireflective (AR) coating - (b) and SEM photographs also show precipitating In solder - (c).**

The temperature map for emitter no. 18 is presented in Fig 5.

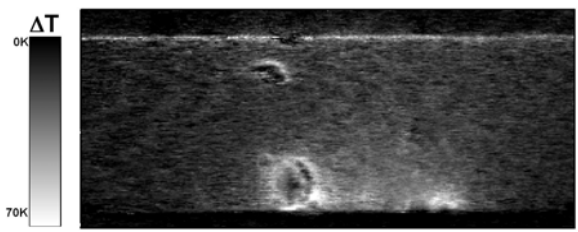

**Fig 5. Temperature map of the emitter no. 8 at I=50A and at the heat sink temperature of 25°C.**

This emitter facet does not exhibit features of drastically damaged coatings as was in the case for emitter no. 8.

However spots of higher temperature in the active region are still observed. In the CCD image captured during measurement dark spots (no lasing) can be observed. Near field image was recorded in order to compare it with the horizontal temperature profile along junction obtained by thermoreflectance spectroscopy. Result of this comparison are presented in Fig. 7.

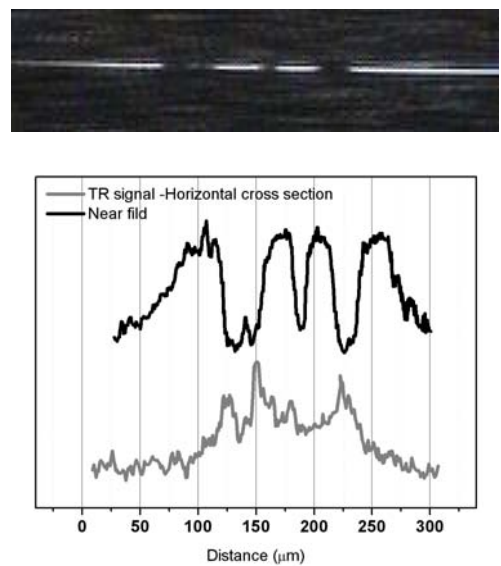

**Fig 7. Comparison of near field and horizontal temperature profile along junction for emitter no. 18.**

The degradation or damage of the facet often leads to appearance of points with lower intensity 'dark spots' in the near field profile. As it can be seen in Fig. 7 a higher value of temperature occurs at the dark spots. In the dark spot regions almost all electric power supplied is converted into heat and also the current density is higher.

The maps recorded for emitters which show no excessive heating exhibit uniform temperature distribution and increase of the overall facet temperature with supply current (see Fig.8).

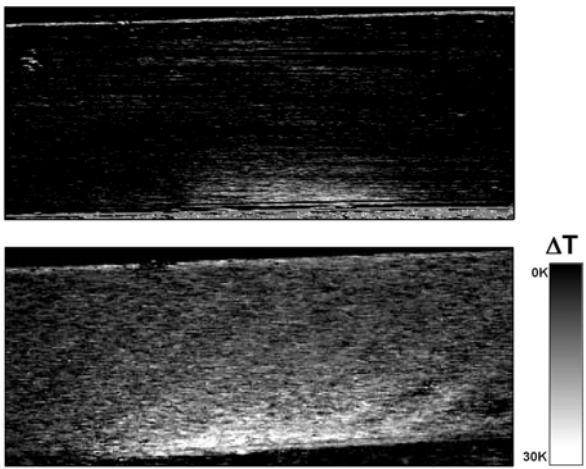

**Fig 8. Temperature maps of the emitter no. 16 at I=30 A (a) and I=50 A (b) at the heat sink temperature of 25°C.**

## 4. CONCLUSION

We have presented a study of the facet heating in high power laser bars. The experimental technique used was spatially resolved thermoreflectance. It allowed for identification of emitter failure mechanisms and gave insight into the heat management of the devices. The observed ageing effects for the individual emitters appeared to be almost exclusively related to facet damage.

We have also observed thermal runaway effects for the laser bar as a whole; i.e., channelling of the supply current by the degraded emitter and consequent further increase of its temperature. The thermoreflectance proved to be method of choice for studying thermal properties of lasers, especially when exact temperature distribution is of interest. The close correlation between TR maps and SEM images of the facet damage additionally validate the method used in the investigations and provide a step forward in quantifying COD effects in high power lasers.